\PassOptionsToPackage{dvipsnames,svgnames,table}{xcolor}

\documentclass[twoside]{aiml22}

\usepackage{aiml22macro}

\usepackage{graphicx}
\usepackage{amsmath}
\usepackage{soul}
\usepackage{url}
\usepackage{hyperref}
\usepackage[utf8]{inputenc}
\usepackage{caption}
\usepackage{amsfonts}
\usepackage{booktabs}
\usepackage{algorithm}
\usepackage{algorithmic}
\usepackage{listings}
\usepackage{textcomp}
\usepackage{multirow}
\usepackage{array}
\usepackage{adjustbox}
\usepackage{balance}
\usepackage{xspace}
\usepackage{tikz}
\usepackage{pgfplots}

\usepackage{main}

\newcolumntype{H}{>{\setbox0=\hbox\bgroup}c<{\egroup}@{}}
\usetikzlibrary{arrows,automata,calc,shapes,decorations,backgrounds,petri,mindmap,fit,positioning}
\usetikzlibrary{calc}

\newcommand{\mi}[1]{\mathit{#1}}

\newcommand{\Coercer}{\mathit{Coercer}}
\newcommand{\Voter}[1][]{\mathit{Voter_{#1}}}
\newcommand{\Selene}{\textbf{\sc Selene}\xspace}
\newcommand{\Domino}{\textbf{Domino DFS}\xspace}
\newcommand{\Uppaal}{\textsc{Uppaal}\xspace}
\newcommand{\Pret}{Pr\^et \`a Voter\xspace}
\newcommand{\wbb}{\text{WBB}}
\newcommand{\tracker}{\mathit{tr}}

\newcommand{\voter}{\mathit{v}}
\newcommand{\voters}{\textit{Voters}}

\newcommand{\vote}{\mathit{Vote}}

\newcommand{\reveal}{\alpha}
\newcommand{\fakereveal}{\alpha'}
\newcommand{\state}{\mathit{g}}
\renewcommand{\model}{\mathit{M}}
\newcommand{\AMAS}{S\xspace}

\newcommand{\fullmodel}{\mathit{IIS}}
\newcommand{\hatPV}{\mathit{PV}}
\newcommand{\evt}{\alpha}								
\newcommand{\evttwo}{\beta}             
\newcommand{\events}{\mathit{Evt}}
\newcommand{\roc}{R} 										
\newcommand{\satisfS}{\models_S}				
\newcommand{\initial}{I}						    
\renewcommand{\lan}[1]{\textbf{#1}\xspace}
\renewcommand{\LTL}{\lan{LTL}}
\renewcommand{\CTL}{\lan{CTL}}

\renewcommand{\ATL}{\lan{ATL}}

\renewcommand{\ATLs}{\ensuremath{\mathbf{ATL^*}}\xspace}
\newcommand{\ATLKs}{\ensuremath{\mathbf{ATL^*K}}\xspace}
\renewcommand{\sATLs}{\ensuremath{\mathbf{sATL^*}}\xspace}
\newcommand{\sATLKs}{\ensuremath{\mathbf{sATL^*K}}\xspace}
\renewcommand{\ATLir}{\ensuremath{\mathbf{ATL_{ir}^*}}\xspace}
\newcommand{\Y}{Y}
\renewcommand{\Y}{}
\renewcommand{\AE}{\textbf{AE}\xspace}
\newcommand{\stacked}[2]{{\Large${#1}\atop{#2}$}}
\newcommand{\agtchoice}{\textit{\scriptsize E}}
\newcommand{\Std}{\textup{Std}\xspace}
\newcommand{\React}{\textup{React}\xspace}
\newcommand{\voterno}{1}
\newcommand{\eps}{{\text{\normalsize$\epsilon$}}}
\newcommand{\IISEps}{IIS^\eps}
\newcommand{\ampleset}{E}
\newcommand{\resultsScale}{0.75}

\definecolor{mygreen}{rgb}{0,0.6,0}
\definecolor{mygray}{rgb}{0.5,0.5,0.5}
\definecolor{mymauve}{rgb}{0.58,0,0.82}

\def\lastname{Kurpiewski, Jamroga, Maśko, Mikulski, Pazderski, Penczek and Sidoruk}

\begin{document}

\begin{frontmatter}
  \title{Verification of Multi-Agent Properties\\in Electronic Voting: A Case Study}
  \author{Damian Kurpiewski$^{1,3}$, Wojciech Jamroga$^{1,2}$, Łukasz Maśko$^{1}$}
			\renewcommand*{\thefootnote}{\fnsymbol{footnote}}
	\author{Łukasz Mikulski$^{3,1}$, Witold Pazderski$^{1}$}
	\author{Wojciech Penczek$^{1}$, and Teofil Sidoruk$^{1,4}$}\footnote[1]{
		The authors acknowledge the support of the National Centre for Research and Development, Poland (NCBR),
		and the Luxembourg National Research Fund (FNR), under the PolLux/FNR-CORE project STV (POLLUX-VII/1/2019).
		W. Penczek and T. Sidoruk acknowledge the support of CNRS/PAS under the project MOSART. }
			\renewcommand*{\thefootnote}{\arabic{footnote}}
  \address{$^1$ Institute of Computer Science, Polish Academy of Sciences\\
		ul. Jana Kazimierza 5, 01-248 Warsaw, Poland}
  \address{$^2$ Interdisciplinary Centre for Security, Reliability, and Trust, SnT, University of Luxembourg\\
		29 Av. John F. Kennedy, 1855 Luxembourg, Luxembourg}
	\address{$^3$ Faculty of Mathematics and Computer Science, Nicolaus Copernicus Univeristy\\
		ul. Chopina 12/18, 87-100 Toruń, Poland}
	\address{$^4$ Faculty of Mathematics and Information Science, Warsaw University of Technology\\
		ul. Koszykowa 75, 00-662 Warsaw, Poland}

  \begin{abstract}
		Formal verification of multi-agent systems is hard, both theoretically and in practice.
		In particular, studies that use a single verification technique typically show limited efficiency, and allow to verify only toy examples.
		Here, we propose some new techniques and combine them with several recently developed ones to see what progress can be achieved for a real-life scenario.
		Namely, we use fixpoint approximation, domination-based strategy search, partial order reduction, and parallelization to verify heterogeneous 
		scalable models of the \Selene e-voting protocol.
		The experimental results show that the combination allows to verify requirements for much more sophisticated models than previously.
  \end{abstract}

 \end{frontmatter}

\section{Introduction}\label{sec:intro}
\emph{Multi-agent systems (MAS)} provide models and methodologies for analysis of systems that feature interaction of multiple autonomous components~\cite{Shoham09MAS,Weiss99mas,Wooldridge02intromas}.
Formal specification and verification of such systems becomes essential due to the dynamic development of AI solutions that enter practical applications~\cite{Akintunde20neuralCTL,Akintunde20neuralATL}.
In particular, it is crucial to assess requirements that refer to \emph{strategic abilities} of agents and their groups, such as the ability of a passenger to leave an autonomous cab (preferably alive), or the inability of an intruder to take remote control of the cab.

\para{Specification and verification of MAS.}
Properties of this kind can be conveniently specified in \emph{modal logics of strategic ability}, of which alternating-time temporal logic \ATLs~\cite{Alur97ATL,Alur02ATL} is probably the most popular.
Logic-based methods for MAS are relatively well studied from the theoretical perspective~\cite{Dastani10MAS,Emerson90temporal,Fagin95knowledge,Jamroga15specificationMAS},
including theories of agents and agency~\cite{Broersen01BOID,Broersen06embedding-atl-stit,Cohen90intention,Rao91BDI,Wooldridge00rational}
semantic issues~\cite{Agotnes04atel,Agotnes15handbook,Dima10communicating,Guelev12stratcontexts,Jamroga04ATEL,Schobbens04ATL}, meta-logical properties~\cite{Bulling14comparing-jaamas,Guelev11atl-distrknowldge}, and the complexity of model checking~\cite{Bulling10verification,Dima11undecidable,Guelev11atl-distrknowldge,Schobbens04ATL,Hoek06practicalmcheck}.
There is even a number of model checking approaches and tools. 
Unfortunately, they only admit temporal properties~\cite{Behrmann04uppaal-tutorial,Dembinski03verics,KacprzakLP04,KacprzakNNPPSWZ08,Kant15ltsmin}, 
deal with the less practical case of perfect information strategies~\cite{Alur00mocha,Chen13prismgames,KacprzakP04,KacprzakP05,KanskiNKPN21}, 
treat imperfect information with limited interest and effectiveness~\cite{Lomuscio17mcmas}, 
or have restricted verification capabilities~\cite{Akintunde20neuralATL,Kurpiewski19stv-demo,Kurpiewski21stv-demo}.
No less importantly, attempts to verify actual requirements on realistic agent systems have been scarce.

In this paper, we combine and extend some of the recent advances in model checking of modal specifications
for MAS~\cite{Jamroga19fixpApprox-aij,Jamroga18por,Kurpiewski19domination}, 
and apply them to see how far we can get with the verification of an existing e-voting protocol.
Anonymous, coercion-resistant, and verifiable e-voting procedures have been proposed and studied 
for over 10 years now~\cite{Chaum05pretavoter,Gardner09e2e-voting,Juels05coercion,Ryan15verifiability}, 
including implementations and their use in real-life elections~\cite{Adida08helios,Chaum09scantegrityII,Culnane15vvote}.
This makes e-voting a great case for testing verification algorithms and tools, being developed for MAS.

\para{Contribution.}
Verification for modal logics of strategic ability is hard, both theoretically and in practice.
Likely, no single technique suffices to deal with it alone.
Here, we try the ``all out'' approach, and combine several techniques, developed recently by our team, to verify properties of the \Selene e-voting protocol~\cite{Ryan16selene}.
We use the algorithms of \emph{fixpoint approximation}~\cite{Jamroga19fixpApprox-aij}, \emph{depth-first} and \emph{domination-based strategy search}~\cite{Kurpiewski19domination}, as well as \emph{partial order reduction}~\cite{Jamroga21paradoxes-kr,Jamroga20POR-JAIR}.
To apply the latter, we extend it to handle strategic-epistemic properties, and prove the correctness of the extension.
We also propose and study a distributed variant of the depth-first and domination-based synthesis.
We evaluate the power of the combined approach on a new model of \Selene, consisting of voters, coercers, and the election infrastructure.
While our model does not yet match the complexity of a real-life election, it goes beyond typically used examples.

\para{Related work.}
Formal verification of voting protocols typically focuses on their cryptographic aspects. The multi-agent and social interaction in the models is limited, and the verification restricts to temporal and bisimulation-based properties.
Examples include the automated analysis of \Selene~\cite{Bruni17selene} and Electryo~\cite{Zollinger20electryo} using the Tamarin prover for security protocols.
Theorem proving in first-order logic was also used to capture some socio-technical factors of Helios in~\cite{Martimiano15ceremony}.
Similarly, the interactive theorem prover Coq for higher-order logic was used in~\cite{Ghale18stv-coq,Haines19verified-verifiers}.

A slightly more detailed multi-agent model of the \Pret protocol was used in~\cite{Jamroga20Pret-Uppaal}, but the verification concerned only simple temporal formulas of \CTL.
In~\cite{Jamroga20natvoting}, strategic abilities in \Pret were considered, but the strategies were hand-crafted rather than synthesized in the verification process.
A preliminary formalization of receipt-freeness and coercion resistance using \ATL was shown in~\cite{Tabatabaei16expressing}, but no verification was proposed.
Perhaps the closest work to the present one was our previous attempt to model and verify \Selene in~\cite{Jamroga18Selene}, but the model used there was extremely simple, and the performance results were underwhelming.

\section{Preliminaries}\label{sec:preliminaries}
We first recall the models of asynchronous interaction in MAS, defined in~\cite{Jamroga21paradoxes-kr}
and inspired by~\cite{lomuscio10partialOrder,Priese83apa-nets}.

\subsection{Models of Asynchronous Interaction}

\begin{definition}[Asynchronous MAS]\label{def:amas}
An \emph{asynchronous multi-agent system (AMAS)} $\AMAS$ consists of $n$ agents $\A = \set{1,\dots,n}$,
each associated with a tuple $A_i =(L_i, \events_i, \roc_i, T_i, \PV_i, V_i)$ including a set
of \emph{local states} $L_i=\{l_i^1, l_i^2,\dots,l_i^{n_i}\}$,
a nonempty finite set of \emph{events} $\events_i=\{\evt_i^1,\evt_i^2,\ldots, \evt_i^{m_i}\}$,
and a \emph{repertoire of choices}
$\roc_i: L_i \to \powerset{\powerset{\events_i}}$.
For each $l_i\in L_i$, $\roc_i(l_i)=\set{\agtchoice_1,\dots,\agtchoice_m}$ is a nonempty list of nonempty choices available to $i$ at $l_i$.
If the agent chooses $\agtchoice_j = \set{\evt_1,\evt_2,\dots}$, then only an event in $\agtchoice_j$ can be executed at $l_i$ within the agent's module.
Moreover, $T_i: L_i \times \events_i \fpart L_i$ is a (partial) \emph{local transition function} such that
$T_i(l_i,\evt)$ is defined iff $\evt\in \bigcup\roc_i(l_i)$.

Agents are endowed with mutually disjoint, finite and possibly empty sets of \emph{local propositions} $\PV_i$,
and their \emph{valuations} $V_i : L_i \then \powerset{\PV_i}$.
$\events = \bigcup_{i \in \A} \events_i$ is the set of all events, and
$Agent(\evt) = \set{i\in\A \mid \evt \in \events_i}$ the set of agents who have access to {event }$\evt$.
$\PV = \bigcup_{i\in\A} \PV_i$ is the set of all propositions.
\end{definition}

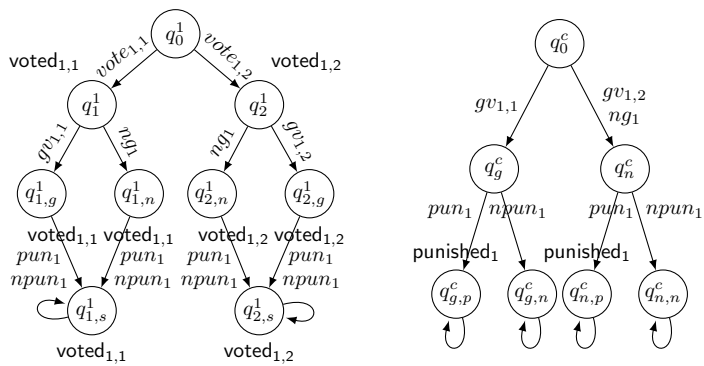
\begin{figure}[t]\centering
\hspace{-0.85cm}
\begin{tabular}{@{}c@{\quad\quad}c@{}}
\begin{tabular}{@{}c@{}}
\begin{tikzpicture}[>=latex,transform shape,scale=0.8]

\tikzstyle{every state}=[fill=none,draw=black,text=black,inner sep=0pt,minimum size=8mm]

\node[state] (s0)	  {$q_0^\voterno$};

\node[state] (s1) [label={100:{$\prop{voted_{\voterno,{1}}}$}}, below left=0.6cm and 0.8cm of s0] 	{$q_{1}^\voterno$};
\node[state] (s2) [label={270:{${}\qquad \prop{voted_{\voterno,{1}}}$}}, below left=0.9cm and 0.25cm of s1]		{$q_{{1},g}^\voterno$};
\node[state] (s3) [label={270:{$\prop{voted_{\voterno,{1}}}$}}, below right=0.9cm and 0.2cm of s1]	  {$q_{{1},n}^\voterno$};
\node[state] (s4) [label={270:{$\prop{voted_{\voterno,{1}}}$}}, below=2.6cm of s1] {$q_{{1},s}^\voterno$};

\node[state] (s1bis) [label={80:{$\prop{voted_{\voterno,{2}}}$}}, below right=0.6cm and 0.8cm of s0] 	{$q_{2}^\voterno$};
\node[state] (s2bis) [label={270:{$\prop{voted_{\voterno,{2}}}$}}, below right=0.9cm and 0.25cm of s1bis]		{$q_{{2},g}^\voterno$};
\node[state] (s3bis) [label={270:{${}\qquad \prop{voted_{\voterno,{2}}}$}}, below left=0.9cm and 0.2cm of s1bis]	  {$q_{{2},n}^\voterno$};
\node[state] (s4bis) [label={[label distance=0.0cm]270:{$\prop{voted_{\voterno,{2}}}$}}, below=2.6cm of s1bis] {$q_{{2},s}^\voterno$};
	
\path[->]
(s0)
  edge node[sloped, anchor=center, above] {$vote_{\voterno,{1}}$}	(s1)
(s1)
  edge node[sloped, anchor=center, above] {$gv_{\voterno,{1}}$} (s2)
  edge node[sloped, anchor=center, above] {$ng_\voterno$} (s3)
(s2)
  edge node [left=-2pt,near end] {\stacked{pun_\voterno}{npun_\voterno}} (s4)
(s3)
  edge node [right=-2pt,near end] {\stacked{pun_\voterno}{npun_\voterno}} (s4)
(s4)
  edge [loop left] node {} (s4)
(s0)
  edge node[sloped, anchor=center, above] {$vote_{\voterno,{2}}$}	(s1bis)
(s1bis)
  edge node[sloped, anchor=center, above] {$gv_{\voterno,{2}}$} (s2bis)
  edge node[sloped, anchor=center, above] {$ng_\voterno$} (s3bis)
(s2bis)
  edge node [right=-2pt,near end] {\stacked{pun_\voterno}{npun_\voterno}} (s4bis)
(s3bis)
  edge node [left=-3pt,near end] {\stacked{pun_\voterno}{npun_\voterno}} (s4bis)
(s4bis)
  edge [loop right] node {} (s4bis);
\end{tikzpicture}
\end{tabular}
 &
\begin{tabular}{@{}c@{}}
\begin{tikzpicture}[>=latex,transform shape,scale=0.8]

\tikzstyle{every state}=[fill=none,draw=black,text=black,inner sep=0pt,minimum size=8mm]

\node[state] (c0) {$q_0^c$};
\node[state] (c1) [below left=1.5cm and 0.5cm of c0] {$q_{g}^c$};
\node[state] (c2) [below right=1.5cm and 0.5cm of c0] {$q_{n}^c$};
\node[state] (c3) [label={90:{$\prop{punished_1}$}}, below left=1.5cm and 0.05cm of c1] {$q_{g,p}^c$};
\node[state] (c4) [below right=1.5cm and 0.05cm of c1] {$q_{g,n}^c$};
\node[state] (c5) [label={90:{$\prop{punished_1}$}}, below left=1.5cm and 0.05cm of c2] {$q_{n,p}^c$};
\node[state] (c6) [below right=1.5cm and 0.05cm of c2] {$q_{n,n}^c$};

\path[->]
(c0)
   edge node [left] {$gv_{\voterno,1}$} (c1)
(c0)
   edge node [right] {\stacked{gv_{\voterno,2}}{ng_\voterno}} (c2)
(c1)
   edge node [left,near start] {$pun_\voterno$} (c3)
(c1)
   edge node [near start] {{}\quad $npun_\voterno$} (c4)
(c2)
   edge node [near start] {$pun_\voterno$} (c5)
(c2)
   edge node [right,near start] {$npun_\voterno$} (c6)
(c3)
   edge [loop below] node {} (c3)
(c4)
   edge [loop below] node {} (c4)
(c5)
   edge [loop below] node {} (c5)
(c6)
   edge [loop below] node {} (c6);
\end{tikzpicture}
\end{tabular}
\end{tabular}
\caption{ASV$_1^2$: agents $\Voter[1]$ (left) and $\Coercer$ (right)}
\label{fig:votingmodel}
\end{figure}

Note that each agent ``owns'' the events affecting its state, but some of the events may be shared with other agents. Such events can only be executed synchronously by all the involved parties. This way, the agent can influence how the states of the other agents evolve.
Moreover, the agent's strategic choices are restricted by its repertoire function.
Assigning sets rather than single events in $\roc_i$, which subsequently determines the type of strategy functions in Section~\ref{sec:atl},
is a deliberate decision, allowing to avoid certain semantic issues that fall outside the scope of this paper.
We refer the reader to~\cite{Jamroga21paradoxes-kr} for the details.

The following example demonstrates a simple AMAS,
while also introducing some key concepts that will be expanded upon in our model of the real-world protocol \Selene (discussed in Section~\ref{sec:specification}).
In particular, there is a \emph{coercer} agent, whose goal is to ensure that the voter(s) select a particular candidate.
To that end, the coercer may threaten them with punishment, e.g. if they refuse to cooperate by not sharing the ballot, or openly defy by voting for another candidate.
\begin{example}[Asynchronous Simple Voting]\label{ex:asv}
Consider a simple voting system ASV$_n^k$ with $n+1$ agents ($n$ voters and $1$ coercer).
Each $\Voter_i$ agent can cast her vote for a candidate $\set{1,\dots,k}$, and decide whether to share her vote receipt with the $\Coercer$ agent.
The coercer can choose to punish the voter or refrain from it.
A graphical representation of the agents for $n=1, k=2$ is shown in Fig.~\ref{fig:votingmodel}.
We assume that the coercer only registers if the voter hands in a receipt for candidate $1$ or not.
The repertoire of the coercer is defined as
$\roc_c(q_0^c)=\set{\set{gv_{\voterno,1},gv_{\voterno,2},ng_{\voterno}}}$ and
$\roc_c(q_g^c)=\roc_c(q_n^c)=\set{\set{pun_\voterno},\set{npun_\voterno}}$,
i.e., the coercer first \emph{receives} the voter's decision regarding the receipt, and then \emph{controls} whether the voter is punished or not.
Analogously, the voter's repertoire is given by: 
$\roc_\voterno(q_0^\voterno)=\set{\set{vote_{\voterno,1}},\set{vote_{\voterno,2}}}$,
$\roc_\voterno(q_j^\voterno)=\set{\set{gv_{\voterno,j}},\set{ng_{\voterno}}}$ for $j=1,2$, 
and
$\roc_\voterno(q_{1,g}^\voterno)=\roc_\voterno(q_{1,n}^\voterno)=\roc_\voterno(q_{2,g}^\voterno)=\roc_\voterno(q_{2,n}^\voterno) = \set{\set{pun_\voterno,npun_\voterno}}$.
\end{example}

Notice that the coercer cannot determine which of the events $gv_{\voterno,1},gv_{\voterno,2},ng_{\voterno}$ will occur; this is entirely under the voter's control. 
This way we model the situation where it is the decision of the voter to show her vote or not.
Similarly, the voter cannot avoid punishment by choosing the strategy allowing only $npun_1$,  
because the choice $\set{npun_\voterno}$ is \emph{not} in the voter's repertoire. She can only execute $\set{pun_\voterno,npun_\voterno}$, and await the decision of the coercer.

The execution semantics is based on interleaving with synchronization on shared events.
Note that for a shared event to be executed, it must be done jointly by all agents who have it in their repertoires.

\begin{definition}[Interleaved interpreted systems]\label{def:canonical-iis}
Let $\AMAS$ be an AMAS with $n$ agents, and let $I\subseteq L_1\times\ldots\times L_n$.
The \emph{full model} $IIS(\AMAS,I)$ extends $\AMAS$ with:\
(i) the set of initial states $I$;\
(ii) the set of global states $\States \subseteq L_1\times\ldots\times L_n$ that collects all the configurations of local states, reachable from $I$ by $T$ (see below);\
(iii) the (partial) \emph{global transition function} $T: \States\times \events \fpart \States$, defined by $T(\state_1,\evt)= \state_2$ iff $T_i(\state_1^i,\evt) = \state^i_2$
for all $i \in Agent(\evt)$ and $\state_1^i = \state^i_2$ for all $i \in \A \setminus Agent(\evt)$;\footnote{
  $\state^i$ denotes agent $i$'s state in $\state=(l_1,\dots,l_n)$, i.e., $\state^i=l_i$. }\
(iv) the \emph{global valuation} of propositions $V: \States \rightarrow 2^{\PV}$, defined as $V(l_1,\dots,l_n) = \bigcup_{i\in\A} V_i(l_i)$.

We will sometimes write $\state_1 \trans \evt \state_2$ instead of $T(\state_1,\evt) = \state_2$.
\end{definition}

\begin{definition}[Enabled events]\label{def:enabled}
Event $\evt \in \events$ is \emph{enabled} at $\state\in \States$ if $\state \trans \evt \state'$ for some $\state' \in \States$,
i.e., $T(\state,\evt)=\state'$.
The set of such events is denoted by $enabled(\state)$.

Let $A=\set{a_1,\dots,a_k} \subseteq\A=\{1,\ldots,n\}$ and $\overrightarrow{\agtchoice}_A = (\agtchoice_{a_1},\dots,\agtchoice_{a_k})$
for some $k \leq n$,
such that $\agtchoice_i\in R_i(\state^i)$ for every $i\in A$.
Event $\evttwo \in \events$ is \emph{enabled by the vector of choices $\overrightarrow{\agtchoice}_A$ at $\state\in \States$} iff,
for every $i\in Agent(\evttwo)\cap A$, we have $\evttwo\in\agtchoice_i$, and
for every $i\in Agent(\evttwo)\setminus A$, it holds that $\evttwo\in \bigcup\roc_i(\state^i)$.
That is, the ``owners'' of $\evttwo$ in $A$ have selected choices that admit $\evttwo$,
while all the other ``owners'' of $\evttwo$ \emph{might} select choices that do the same.
We denote the set of such events by $enabled(\state,\overrightarrow{\agtchoice}_A)$.
Clearly, $enabled(\state,\overrightarrow{\agtchoice}_A) \subseteq enabled(\state)$.
\end{definition}

Some combinations of choices enable no events.
To account for this, the models of AMAS are augmented with ``silent'' $\epsilon$-loops, added when no ``real'' event can occur.
\begin{definition}[Undeadlocked IIS]\label{def:undeadlockedIIS}
Let $\AMAS$ be an AMAS, and assume that no agent in \AMAS has $\epsilon$ in its alphabet of events.
The \emph{undeadlocked model of \AMAS}, denoted $\IISEps(\AMAS,\initial)$, extends the model $IIS(\AMAS,\initial)$ as follows:
\begin{itemize}
\item $\events_{\IISEps(\AMAS,\initial)} = \events_{IIS(\AMAS,\initial)} \cup \set{\epsilon}$, where $Agent(\epsilon) = \emptyset$;\
\item For each $\state\in\States$, we add the transition $\state \trans\epsilon \state$
  iff there is a selection of all agents' choices $\overrightarrow{\agtchoice}_{\A} = (\agtchoice_{1},\dots,\agtchoice_{n})$,
  such that $\agtchoice_i\in \roc_i(g^i)$ and $enabled_{IIS(\AMAS,\initial)}(\state,\overrightarrow{\agtchoice}_{\A}) = \emptyset$.
  Then, for every $A\subseteq\A$, we also fix
  $enabled_{\IISEps(\AMAS,I)}(\state,\overrightarrow{\agtchoice}_A) = enabled_{\IISEps(\AMAS,I)}(\state,\overrightarrow{\agtchoice}_A) \cup \set{\epsilon}$.

  In other words, an $\epsilon$-loop is enabled whenever $\agtchoice_A$ allows the grand coalition to collectively block the execution of any ``real'' event.
\end{itemize}
\end{definition}

We use the term \emph{model} to refer to subgraphs of $IIS(\AMAS,I)$ as well as $\IISEps(\AMAS,I)$.

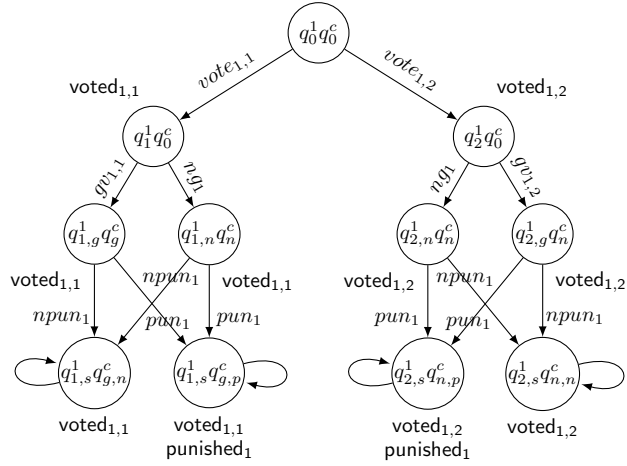
\begin{figure}[t]\centering
\hspace{-0.6cm}
\begin{tikzpicture}[>=latex,transform shape,scale=0.8]

\tikzstyle{every state}=[fill=none,draw=black,text=black,inner sep=0pt,minimum size=10mm]

\node[state] (s0)	  {$q_0^\voterno q_0^c$};

\node[state] (s1) [label={100:{$\prop{voted_{\voterno,{1}}}$}}, below left=1cm and 2cm of s0] 	{$q_{1}^\voterno q_0^c$};
\node[state] (s2) [label={260:{${}\qquad \prop{voted_{\voterno,{1}}}$}}, below left=0.9cm and 0.25cm of s1]		{$q_{{1},g}^\voterno q_{g}^c$};
\node[state] (s3) [label={280:{$\prop{voted_{\voterno,{1}}}$}}, below right=0.9cm and 0.2cm of s1]	  {$q_{{1},n}^\voterno q_{n}^c$};
\node[state] (s4) [label={270:{$\prop{voted_{\voterno,{1}}}$}}, below=1.2cm of s2] {$q_{{1},s}^\voterno q_{g,n}^c$};
\node[state] (s5) [label={270:{\stacked{\prop{voted_{\voterno,{1}}}}{\prop{punished_{\voterno}}}}}, below=1.2cm of s3] {$q_{{1},s}^\voterno q_{g,p}^c$};

\node[state] (s1bis) [label={80:{$\prop{voted_{\voterno,{2}}}$}}, below right=1cm and 2cm of s0] 	{$q_{2}^\voterno q_0^c$};
\node[state] (s2bis) [label={280:{$\prop{voted_{\voterno,{2}}}$}}, below right=0.9cm and 0.25cm of s1bis]		{$q_{{2},g}^\voterno q_{n}^c$};
\node[state] (s3bis) [label={260:{${}\qquad \prop{voted_{\voterno,{2}}}$}}, below left=0.9cm and 0.2cm of s1bis]	  {$q_{{2},n}^\voterno q_{n}^c$};
\node[state] (s4bis) [label={270:{$\prop{voted_{\voterno,{2}}}$}}, below=1.2cm of s2bis] {$q_{{2},s}^\voterno q_{n,n}^c$};
\node[state] (s5bis) [label={270:{\stacked{\prop{voted_{\voterno,{2}}}}{\prop{punished_{\voterno}}}}}, below=1.2cm of s3bis] {$q_{{2},s}^\voterno q_{n,p}^c$};
	
\path[->]
(s0)
  edge node[sloped, anchor=center, above] {$vote_{\voterno,{1}}$}	(s1)
(s1)
  edge node[sloped, anchor=center, above] {$gv_{\voterno,{1}}$} (s2)
  edge node[sloped, anchor=center, above] {$ng_\voterno$} (s3)
(s2)
  edge node [left=-2pt,near end] {$npun_\voterno$} (s4)
  edge node [near end] {$pun_\voterno$} (s5)
(s3)
  edge node [near start] {$npun_\voterno$} (s4)
  edge node [right=-0pt,near end] {$pun_\voterno$} (s5)
(s4)
  edge [loop left] node {} (s4)
(s5)
  edge [loop right] node {} (s5)
(s0)
  edge node[sloped, anchor=center, above] {$vote_{\voterno,{2}}$}	(s1bis)
(s1bis)
  edge node[sloped, anchor=center, above] {$gv_{\voterno,{2}}$} (s2bis)
  edge node[sloped, anchor=center, above] {$ng_\voterno$} (s3bis)
(s2bis)
  edge node [right=-2pt,near end] {$npun_\voterno$} (s4bis)
  edge node [near end] {$pun_\voterno$} (s5bis)
(s3bis)
  edge node [near start] {$npun_\voterno$} (s4bis)
  edge node [left=-0pt,near end] {$pun_\voterno$} (s5bis)
(s4bis)
  edge [loop right] node {} (s4bis)
(s5bis)
  edge [loop left] node {} (s5bis);
\end{tikzpicture}
\caption{The undeadlocked model $\IISEps(ASV_1^2)$}
\label{fig:votingmodel-iis}
\end{figure}

\begin{example}\label{ex:asv-model}
The undeadlocked model of ASV$_1^2$ is shown in Figure~\ref{fig:votingmodel-iis}.
Note that it contains no $\eps$-transitions, since no choices of the voter and the coercer can cause a deadlock.
\end{example}

\subsection{Reasoning About Strategies and Knowledge}\label{sec:atl}

Let $\PV$ be a set of propositions
and $\A$ the set of all agents.
The syntax of \emph{alternating-time logic} \ATLs~\cite{Alur02ATL,Schobbens04ATL} is given by:
\begin{center}
$\varphi::= \prop{p} \mid \neg \varphi \mid \varphi\wedge\varphi \mid \coop{A}\gamma$, \hspace{1cm}
$\gamma::=\varphi \mid \neg\gamma \mid \gamma\land\gamma \mid \Next\gamma \mid \gamma\Until\gamma$, \\
\end{center}
where $\propp \in \PV$, $A \subseteq \A$, $\Next$ stands for ``next'',
$\Until$ for ``until'',
and $\coop{A}\gamma$ for ``agent coalition $A$ has a strategy to enforce $\gamma$''.
Temporal operators $\Sometm$ (``eventually'') and $\Always$ (``always''), Boolean connectives, and constants are defined as usual.

\begin{example}\label{ex:coerc-feas}
\emph{Coercion feasibility} against voter $i$ can be expressed by formula
$\coop{\Coercer}\Sometm\prop{punished_i}$ (the coercer can ensure that the voter is eventually punished).
\end{example}

\para{Strategic ability of agents.} Following~\cite{Jamroga21paradoxes-kr}, a \emph{positional imperfect information strategy (\ir-strategy) for agent $i$}
is defined by a function $\strat_i \colon L_i \to \powerset{\events_i}$,
such that $\strat_i(l) \in \roc_i(l)$ for each $l \in L_i$.
Note that $\strat_i$ is uniform by construction, as it is based on local, and not global states.
The set of such strategies is denoted by $\Sigma_i^{\ir}$.
Joint strategies $\Sigma_A^{\ir}$ for $A=\set{a_1,\dots,a_k} \subseteq\A$ are defined as usual,
i.e., as tuples of strategies $\strat_i$, one for each agent $i \in A$.
By $\strat_A(\state) = (\strat_{a_1}(\state),\dots,\strat_{a_k}(\state))$, we denote the joint choice of coalition $A$ at global state $\state$.
An infinite sequence of global states and events $\seq = \state_0 \evt_0 \state_1 \evt_1 \state_2\dots$ is
called a \emph{path} if $\state_j \trans {\evt_j} \state_{j+1}$ for every $j \geq 0$.
The set of all paths in model $M$ starting at state $\state$ is denoted by $\Pi_M(\state)$.

\begin{definition}[Standard outcome]\label{def:outcome}
Let $A \subseteq\A$.
The \emph{standard outcome} of strategy $\strat_A\in\Sigma_A^\ir$ in state $\state$ of model $\model$
is the set $\outcome_M^\Std(\state,\strat_A) \subseteq \Pi_M(\state)$ such that
$\seq = \state_0 \evt_0 \state_1 \evt_1 \dots \in \outcome_M(\state,\strat_A)$
iff $\state_0 = \state$,
and for each $m \geq 0$ we have that $\evt_m \in enabled_M(\state_m,\strat_A(\state_m))$.
\end{definition}

\begin{definition}[Reactive outcome]\label{def:reactive-outcome}
The \emph{reactive outcome}
is the set $\outcome_M^\React(\state,\strat_A) \subseteq \outcome_M^\Std(\state,\strat_A)$ such that
$\seq = \state_0 \evt_0 \state_1 \evt_1 \dots \in \outcome_M^\React(\state,\strat_A)$
iff $\evt_m = \epsilon$ implies $enabled_M(\state_m,\strat_A(\state_m)) = \set{\epsilon}$.
\end{definition}
Intuitively, the standard outcome collects all the paths where agents in $A$ follow $\strat_A$, while the others freely choose from their repertoires.
The reactive outcome includes only those outcome paths where the opponents cannot miscoordinate on shared events.
Let $x\in\set{\Std,\React}$.
The \ir-semantics of $\coop{A}\gamma$ in asynchronous MAS~\cite{Alur02ATL,Schobbens04ATL,Jamroga21paradoxes-kr} is defined by the clause:
\begin{description2}
\setlength\itemindent{-0.25cm}
\item[{$\model,\state \satisf^x \coop{A}\gamma$}] iff there is a strategy $\strat_A\!\in\!\Sigma_A^\ir$
such that
for all $\seq\!\in\! \outcome_\model^x(\state,\strat_A)$
we have $\model,\seq\! \satisf^x\! \gamma$.
\end{description2}

\para{Adding knowledge operators.}
The following relations capture the notion of indistinguishability between states, needed to define semantics for the epistemic modality.

\begin{definition}[Indistinguishable states]\label{def:indistinguishable}
For each $i \in \A$, the relation $\sim_i = \{(\state,\state') \in St \times St \mid \state^i = \state'^i \}$
denotes that states $\state, \state'$ are \textit{indistinguishable} for agent $i$.
The relation $\sim_J = \bigcap_{j \in J} \sim_j$ extends it to the distributed knowledge of a group of agents $J \subseteq \A$.
\end{definition}

By \ATLKs, we denote the extension of \ATLs with knowledge operators $\K_i$
where $\K_i\psi$ means ``agent $i$ knows that $\psi$''.
Note that temporal and strategic operators cannot be nested inside $\psi$.
The semantics of $\K_i\psi$ can be defined by the clause~\cite{Fagin95knowledge,Hoek02ATEL}:
\begin{description}
\item[{$\model,\state \satisf[\Y] \K_i \psi$}] iff $\model,\state' \satisf[\Y] \psi$ for every $\state'$
such that $\state' \sim_i \state$.
\end{description}

In e-voting, epistemic properties arise due to information exchanged by cryptographic protocols, but also published in plaintext, e.g., on the Web Bulletin Board.

\para{Subjective ability.}
The asynchronous semantics of $\coop{A}\gamma$ in~\cite{Jamroga21paradoxes-kr} is based on the notion of ``objective'' ability, i.e., it suffices that
there exists a strategy $\strat_A$ which enforces $\gamma$ on outcome paths from the objective starting point $\state$ of the model.
The more popular notion of ``subjective'' ability requires that $\strat_A$ succeeds on all outcome paths from the states that $A$ might consider as possible starting points, cf.~\cite{Bulling14comparing-jaamas} for an in-depth discussion.
The ``subjective'' semantics of strategic operators can be defined as: 
\begin{description}
\item[{$\model,\state \satisfS^x \coop{A}\gamma$}] iff there is a strategy $\strat_A\in\Sigma_A^\ir$ such that, for each $\seq\in \bigcup_{i\in A}\bigcup_{\state' \sim_i \state} \outcome_\model^x(\state',\strat_A)$, we have $\model,\seq \satisfS^x \gamma$.
\end{description}

\begin{example}
Let $M = \IISEps(ASV_n^k,\set{\state_0})$ where $\state_0=(q_0^1,\dots,q_0^n,q_0^c)$. That is, $M$ is the undeadlocked model of the AMAS in Example~\ref{ex:asv} with $\state_0$ as its sole initial state.
Note that the $\Std$ and $\React$ semantics coincide on $M$, as it includes no $\eps$-transitions.
Clearly, $M,\state_0 \models \coop{\Coercer}\Sometm\prop{punished_i}$ for both the objective and subjective semantics.
On the other hand, the stronger requirement $\coop{\Coercer}\Sometm K_i\prop{punished_i}$ does not hold in $(M,\state_0)$.
\end{example}

In this paper, we focus on the (more popular) subjective semantics.
Moreover, we only use formulas with no next step operators $\Next$ and no nested strategic modalities, which is essential for the application of partial-order reduction.
The corresponding ``simple'' subset of \ATLs (resp. \ATLKs) is denoted by \sATLs (resp. \sATLKs).
The restriction is less prohibitive than it seems at a glance.
First, the $\Next$ operator is of little value for asynchronous systems.
Secondly, nested strategic modalities would only allow us to express an agent’s ability to endow another agent with ability (or deprive the other agent of ability).
Such properties are sometimes interesting, e.g., one may want to require that $\coop{\Voter[\voterno]}\Always \lnot \coop{\Coercer}\Sometm \prop{punished_\voterno}$ (the voter can keep the coercer unable to punish the voter).
Still, simpler properties like $\coop{\Voter[\voterno]}\Always \lnot \prop{punished_\voterno}$ and $\coop{\Voter[\voterno]}\Always \bigwedge_{j=1,\dots,k} \lnot K_\Coercer \prop{voted_{\voterno,j}}$ are usually of more immediate interest.

Finally, we remark that, for subjective ability, epistemic operators are definable with strategic operators, since $K_i\varphi \equiv \coop{i}\bot\Until\varphi$. Moreover, $M,g \models \coop{i}\gamma$ always implies $M,g \models K_i\coop{i}\gamma$ in the subjective semantics.

\section{\mbox{How to Specify Voters and Coercers}}\label{sec:specification}
To keep the paper self-contained, in this section we provide a short description of the \Selene
voting protocol and its formal specification.

\subsection{Short Description of \Selene}\label{sec:selene}

\Selene~\cite{Ryan16selene} is an electronic voting protocol aimed to provide an effective mechanism for \emph{voter verifiability} and \emph{coercion resistance}.
On the one hand, the voter receives a piece of evidence that allows to check if her vote has been registered correctly.
On the other hand, she can present a fake vote evidence to the coercer, thus convincing him that she voted according to the coercer's request.

The protocol proceeds as follows.
Before the election, the Election Authority ($EA$) sets up the system, generating the election keys used for the encryption and decryption of the votes and preparing the vote trackers, one per voter.
The trackers are then encrypted and mixed to break any link between the voter and her tracker, and published on the Web Bulletin Board (\wbb).

In the voting phase, each voter fills in, encrypts, and signs her vote, and sends it to the system.
After several intermediate steps, a pair $(\vote_{\voter}, \tracker_{\voter})$, consisting of the decrypted ballot and the tracker of $\voter$, is published on the \wbb\ for each $\voter\in\voters$.
At this stage, no voters know their trackers. All the votes are presented in plaintext on the \wbb. Thus, the tally of the election is open to a public audit.

At the final stage, the voters receive their trackers by an independent channel (e.g., sms). If the voter has not been coerced, then she requests the special term $\reveal_{\voter}$, which allows for obtaining the correct tracker $\tracker_\voter$.
If she was coerced to fill her ballot in a certain way, she sends a description of the requested vote to the election server.
After such a request, a fake term $\fakereveal_{\voter}$ is sent to the voter, which can be presented to the coercer. The $\fakereveal_{\voter}$ token, together with
the public commitment of the voter, reveals a tracker pointing out to a vote compatible with the coercer's demand.

\Selene uses the ElGamal key encryption scheme, and relies on multiplicative homomorphism and non-interactive zero-knowledge proofs of knowledge that accompany all the transformations of data presented in \wbb. In this work we abstract away from the cryptography and focus on the interaction between the involved agents.

\subsection{Asynchronous MAS for \Selene}

In our model of \Selene, we concentrate on the non-cryptographic interaction between the agents.
In this sense, the present model is similar to~\cite{Jamroga18Selene}, with three important differences.

First, the new modeling is fully modular and scalable. To this end, we use AMAS as input (rather than concurrent game structures or concurrent interpreted systems).
Secondly, we have defined a flexible specification language for AMAS, based on asynchronous agent templates similar to those of \Uppaal~\cite{Behrmann04uppaal-tutorial}, and implemented an interpreter for it.
Besides modularity and scalability, this allowed us to adapt and use partial order reduction for our models.
Thirdly, our specifications include much more details of the \Selene procedure than the skeletal model in~\cite{Jamroga18Selene}.

\subsection{Agents}

There are 4 templates for agents in the modeling: the Election Authority (EA) that handles the generation and distribution of trackers and provides the Web Bulletin Board for voters and coercers, the Coercer, the standard Voter, and the Coerced Voter interacting with a coercer.
The templates are written in a simple specification language created to provide the input for our algorithms.
Each template consists of the name of the agent, the number of instances of that template in the model (e.g., the number of voters), the initial state, the list of transitions, and the agent's repertoire of choices (called ``PROTOCOL'' here).
The Coerced Voter and Coercer templates for an election with 2 candidates are shown in Figures~\ref{fig:voterCode} and~\ref{fig:coercerCode}.
The full specification is available at \href{https://github.com/blackbat13/stv/blob/master/models/Selene.txt}{github.com/blackbat13/stv/blob/master/models/Selene.txt}.

A template always begins with the keyword $Agent$, the agent's name, and the number of instances.
The next line specifies the initial state, followed by the list of local transitions.
Each transition starts with an event name, optionally preceded by the keyword $\mi{shared}$ if the event is shared with another agent.
Then, the source and the target states are given.
Optionally, the transition specification can also include a precondition (in the form of a simple Boolean formula) and/or a postcondition (via a list of updates specifying propositions and their new values).
These conditions are only a technical shortcut that allows us to write clearer and shorter specifications.
The keyword $\mi{aID}$ represents the ID of the current agent and is automatically replaced when preparing local models of agents. 
For example, the template VoterC[2] would produce two agents, $\Voter[C1]$ and $\Voter[C2]$.
Then, each instance of the keyword $\mi{aID}$ will be replaced with $\Voter[C1]$ for the first coerced voter, and with $\Voter[C2]$ for the second one.

\begin{figure}[!t]
  \begin{lstlisting}
    Agent VoterC[1]:
    init start
    shared coerce1_aID: start -> coerced [aID_required=1]
    shared coerce2_aID: start -> coerced [aID_required=2]
    select_vote1: coerced -> prepared [aID_vote=1, aID_prep_vote=1]
    select_vote2: coerced -> prepared [aID_vote=2, aID_prep_vote=2]
    shared is_ready: prepared -> ready
    shared start_voting: ready -> voting
    shared aID_vote: voting -> vote [Coercer1_aID_vote=?aID_vote, Coercer1_aID_revote=?aID_revote]
    shared send_vote_aID: vote -> send
    revote_vote_1: send -[aID_revote==1]> voting [aID_vote=?aID_required, aID_revote=2]
    skip_revote_1: send -[aID_revote==1]> votingf
    revote_vote_2: send -[aID_revote==2]> voting [aID_vote=?aID_required, aID_revote=3]
    skip_revote_2: send -[aID_revote==2]> votingf
    final_vote: send -[aID_revote==3]> votingf [aID_vote=?aID_prep_vote]
    skip_final: send -[aID_revote==3]> votingf
    shared send_fvote_aID: votingf -> sendf
    shared finish_voting: sendf -> finish
    shared send_tracker_aID: finish -> tracker
    shared finish_sending_trackers: tracker -> trackers_sent
    shared give1_aID: trackers_sent -> interact [Coercer1_aID_tracker=1]
    shared give2_aID: trackers_sent -> interact [Coercer1_aID_tracker=2]
    shared not_give_aID: trackers_sent -> interact [Coercer1_aID_tracker=0]
    shared punish_aID: interact -> ckeck [aID_punish=true]
    shared not_punish_aID: interact -> check [aID_punish=false]
    shared check_tracker1_aID: check -> end
    shared check_tracker2_aID: check -> end
    PROTOCOL: [[coerce1_aID, coerce2_aID], [punish, not_punish]]
    \end{lstlisting}
\vspace{-1.5cm}\caption{Voter template}\label{fig:voterCode}
\end{figure}

\begin{figure}[!t]
  \begin{lstlisting}
    Agent Coercer[1]:
    init coerce
    shared coerce1_VoterC1: coerce -> coerce [aID_VoterC1_required=1]
    shared coerce2_VoterC1: coerce -> coerce [aID_VoterC1_required=2]
    shared start_voting: coerce -> voting
    shared VoterC1_vote: voting -> voting
    shared finish_voting: voting -> finish
    shared finish_sending_trackers: finish -> trackers_sent
    shared give1_VoterC1: trackers_sent -> trackers_sent
    shared give2_VoterC1: trackers_sent -> trackers_sent
    shared not_give_VoterC1: trackers_sent -> trackers_sent
    to_check: trackers_sent -> check
    shared check_tracker1_Coercer1: check -> check
    shared check_tracker2_Coercer1: check -> check
    to_interact: check -> interact
    shared punish_VoterC1: interact -> interact
    shared not_punish_VoterC1: interact -> interact
    finish: interact -> end [aID_finish=1]
    PROTOCOL: [[give1_VoterC1, give2_VoterC1, not_give_VoterC1]]
    \end{lstlisting}
\vspace{-1.5cm}\caption{Coercer template}\label{fig:coercerCode}
\end{figure}

Consider the template in Figure~\ref{fig:voterCode}.
The first step of the coerced voter is her interaction with the coercer who can request the vote for a particular candidate (events $\mi{coerce1\_VoterC1}$ and $\mi{coerce2\_VoterC1}$).
The next step is the event $\mi{start\_voting}$ synchronized with the $EA$.
When the election has begun, the voter can create her commitment, fill in the vote, encrypt it, and send it to the EA.
After that, she waits for the publication of votes, which is controlled by the EA.
When the votes are published on the WBB, the voter can decide to compute the false alpha term and the false tracker using publicly available data, or she can just wait for her real tracker.
The last few steps for the VoterC agent consist of checking the WBB, verifying the vote, and interacting with the coercer again.
The voter can show one of her trackers (the false one or the real one) to the coercer, who then either punishes her or does not.

Additionally, the voter can do \emph{revoting}, i.e., cast her vote multiple times, which is a well-known technique to counter in-house coercion by family members.

\subsection{Specification of Properties}\label{sec:properties}

To specify interesting properties of the voting system, we use simple formulas of \ATLKs.
In the experiments, we will concentrate on the property of \emph{coercion-vulnerability}, using the following formula:
\begin{multline*}
\varphi_{vuln,i,k} \equiv\\ \coop{\Coercer}\Always ((\prop{end} \land \prop{revote_{v1}=k} \land \prop{voted_{v1}=i}) \then \K_{Coercer}\prop{voted_{v1}=i})
\end{multline*}

Formula $\varphi_{vuln,i}$ says that the coercer has a strategy so that, at the end of the election, if the voter has effectively voted for candidate $i$, the coercer knows about it.
This can be seen as the opposite of \emph{receipt-freeness} and \emph{coercion-resistance} formalizations in~\cite{Juels05coercion,Tabatabaei16expressing}.
Note that the formula is parameterized by the name $i$ of the preferred candidate of the coercer, as well as the number $k$ of revoting rounds that the coercer is able to observe and learn the value of the cast vote.
Proposition $\prop{revote_{v1}}$ corresponds to the number of revoting rounds performed by the first voter.

We will use the above models and formulas in our verification experiments in Section~\ref{sec:experiments}.

\subsection{Model definition}

Model file consists of definitions of agents templates and various properties. 
Empty lines and Lines starting with \% are comments lines and are ignored by the parser.
Properties that can be defined in the model file are described below.
Each proposition starts with uppercase property name followed by a colon.

\para{PERSISTENT}
List of propositions names that should be considered as persistent when generating the model.
Persistent proposition retains its value after it was set.
Non-persistent propositions are present only in the state that is the result of the transition that created this proposition.

\para{REDUCTION}
List of propositions names used for the partial-order reduction method.

\para{FORMULA}
\ATL formula to be verified.

\para{SHOW\_EPISTEMIC}
Boolean value used only in the graphical interface.
If set to $true$ links of the epistemic relations will be displayed in the model view.

\subsection{Agent definition}

Agent definition consists of a set of non-empty lines specifying the local model of the agent.
Each agent definition is a template that can be used to generate multiple agents of a given type.
Individual definitions are explained below.

\para{Header}
The first line should specify the name of the agent template and number of agents of this type, given in the following format: \texttt{Agent AgentName[count]:}.
For example, in case of \texttt{Agent Voter[3]:} three agents will be generated: \emph{Voter1}, \emph{Voter2} and \emph{Voter3}.

\para{init}
First line of the agent specification, name of the initial state.

\para{Local transition}
Transition is defined in format: \\ \texttt{actionName: state1 -[preCondition]> state2 [propositions]}. \\
Propositions are given in a form of a comma-separated list of variable definitions, i.e. \texttt{[prop1=true, prop2=false, prop3=2]}.
Precondition can be any boolean formula that can be evaluated in Python.

\para{Shared transition}
Shared transitions are defined similarly to local transitions, but are prefixed with \emph{shared} keyword.

\para{Dynamic names}
In order to simplify the specification dynamic names can be used.
Every occurence of the keyword \emph{aID} in the agent template specification will be replaced with the name of the computed agent.

\section{Verification Algorithms}\label{sec:verification}
Model checking strategic ability under imperfect information is hard, both theoretically and in practice.
In this section, we present two recently developed techniques that we use to tackle the high complexity of verification: fixpoint approximation and dominance-based depth-first strategy search (Sections~\ref{sec:fixpoint} and~\ref{sec:domino}).
We also propose a novel approach based on distributed strategy synthesis for \sATLKs (Section~\ref{sec:parallel}).

\subsection{Verification by Fixpoint Approximation}\label{sec:fixpoint}

The main idea of the fixpoint approximations method presented in~\cite{Jamroga19fixpApprox-aij} is that sometimes, instead of the exact model-checking, it suffices to provide a lower and an upper bound for the output. In particular, given a formula $\phi$, we construct two translations $tr_L(\phi)$ and $tr_U(\phi)$, such that $tr_L(\phi)\Rightarrow \phi \Rightarrow tr_U(\phi)$. In other words, if the lower bound translation $tr_L(\phi)$ evaluates to true, then the original formula $\phi$ must hold in the model. Similarly, if the upper bound translation $tr_U(\phi)$ is verified as false, then $\phi$ is also false.

The approximations are built on fixpoint-definable properties. For the upper bound translation we just compute the given formula under the perfect information assumption. For the lower bound we rely on translations that map the formula of \ATLir to an appropriate variant of alternating $\mu$-calculus, see~\cite{Jamroga19fixpApprox-aij} for more details.

\subsection{Depth-First Strategy Search}

Further, we have implemented a depth-first search algorithm for strategy synthesis.
The typical recursive DFS-based approach needed to be adapted due to the presence of epistemic classes and the nondeterministic outcome of the coalition's actions.
In case of nondeterminism due to multiple possible transitions,
it is possible that decisions taken in one of the branches
may determine some choices in the other branches. If such a decision leads to a locally winning strategy, it may need to be changed since decisions in nodes being members of the same epistemic class can influence the outcome of transitions in other branches.
The proposed algorithm allows to backtrack and change locally winning strategy when no winning strategy is found in another branch.

Clearly, even for relatively small models with hundreds of states, the space of strategies is too big to perform a full search.
To address this, we used our DFS algorithm as the backbone for implementing more refined methods, namely \Domino and two variants of parallel model checking, see Sections~\ref{sec:domino} and~\ref{sec:parallel}.

\subsection{Domination-Based Strategy Search}\label{sec:domino}

The strategy space grows exponentially with respect to the number of states and transitions in the model (on top of the state space explosion).
However, in reality it often suffices to only check a subset of possible strategies. This idea was used in the \Domino algorithm~\cite{Kurpiewski19domination}. In that method, a notion of strategy dominance was proposed, according to which the dominated strategies are omitted during the search. Although in the worst case scenario all the strategies in the model need to be checked, the experimental results in~\cite{Kurpiewski19domination} showed that in some cases the method achieves significant improvement in performance.

\subsection{Parallel Implementation}\label{sec:parallel}

The verification process itself has exponential nature with complexity of $O(k^n)$ for a single agent (where $k$ is the number of possible actions and $n$ is the number of states in the model). To reduce the search time, parallel computation can be used. As a basis and a reference, we used a sequential version of the recursive DFS-based strategy search. The biggest advantage of the recursive approach is its ability to perform multiple, independent searches at a time. Such a solution may be very effective for distributed computing architectures, when the search space can be divided independently and distributed between computing nodes with separate memory spaces. The downside is that all the nodes obtain copies of the same model, but then they all may compute independently looking for different solutions.

We have defined and implemented two different approaches to parallelizing depth-first strategy search.

\paragraph{\mbox{Distributed Strategy Search: Simple Branching.}}\label{sec:distr-simple}

The simple approach tries to concurrently execute a number of instances of sequential search, but assuming different strategy prefixes (i.e., such sequences of actions starting at a starting state, which correspond to deterministic paths in a graph -- except for the last one, where there may be a branch). To achieve this, the algorithm first executes a breadth-first search to determine a set of potential different strategy prefixes up to a number equal to a value given as parameter.
Due to the way in which the prefixes are selected, they correspond to single paths and they can be identified with single states. The algorithm tries to expand the prefixes by performing a BFS graph traversal.

Similarly to most BFS-based algorithms, a queue of pending prefixes is used. Initially, this queue contains an empty prefix denoting only one process to be executed. In each step of the algorithm, the first pending prefix is selected. The algorithm tries to expand this prefix by adding actions available in a state to which this prefix leads. For every deterministic action in such state, a new prefix is generated by appending this action to the current prefix. So obtained new prefix is added to the pending queue for further expansion. If the action is nondeterministic, the prefix obtained by appending this action to the current prefix is added to the set of resulting prefixes and is not expanded any more. The current prefix is then discarded. The loop stops when either there are no more pending prefixes or the number of prefixes in both the result set and the pending queue exceeds the given parameter. In the latter case, all the pending prefixes are copied to the resulting set.

After a set of prefixes is determined, the main process spawns a number of children processes. Each subprocess tries to find a winning strategy using the sequential algorithm, assuming that the strategy starts with actions from its assigned prefix. The main process waits for the results from its children; if any of them reports that a winning solution has been found, all the other children are terminated and the search ends.

\paragraph{Distributed Strategy Search: Flexible Version.}\label{sec:distr-flex}

Our second approach uses concurrent execution to examine parallel branches of a single strategy in a flexible way. To this end, it directly expands the single--threaded DFS method.
In this approach, parallel threads cooperate in building the same subgraph of the model that corresponds to the outcome of the candidate strategy. The initial master thread spawns new worker threads whenever it finds a state in which selection of the same action leads to multiple states. A sequential algorithm in such case would
try to recursively build a substrategy considering all the possible targets one by one. In the concurrent version, an additional worker thread is created for each possible outcome of the same action.

In order to ease the synchronization between threads, only the main thread is allowed to spawn new threads due to parallel branches. It is also assumed that only one parallel thread is allowed to select actions for states from any {epistemic class}. If any worker thread reaches a state in such a class, it must wait until the master thread fixes the action for this class. Synchronization is also needed when one thread meets a node which is already examined by another thread -- since the nodes cooperate, it is not necessary to redo the same work, therefore the second node is suspended until the node is checked by the first one. The main thread is allowed to intercept the work already performed by a worker thread.

\section{Taming State Space Explosion}\label{sec:por}
The main obstacle in model checking of MAS logics is the prohibitive complexity of models, in particular due to the state-space explosion.
Partial order reduction (POR) is a well-known technique for state space reduction,
dating back more than three decades \cite{GodefroidPOR,PeledPOR,ValmariPOR}.
The idea is to restrict the set of all enabled transitions
to a representative (i.e. provably sufficient) subset, based on some underlying notion of equivalence.
Crucially, this occurs while generating the unfolding of the system, so the full model,
which may be far too large, is never created.

POR has been defined for variants of \LTL and \CTL, including temporal-epistemic logics~\cite{lomuscio10partialOrder}.
Furthermore, the reduction for \LTL was recently adapted (notably, at the same computational cost)
to the much more expressive logic \sATLs~\cite{Jamroga18por,Jamroga20POR-JAIR},
allowing to leverage the existing algorithms and tools for the verification of strategic abilities.
In~\cite{Jamroga21paradoxes-kr}, POR for \sATLs was shown to still work under an improved execution semantics for AMAS,
which we also adopt in this paper.
Note, however, that our formalization of coercion-vulnerability in Section~\ref{sec:properties}
is a \emph{strategic-epistemic property}, and hence those results
do not cover this case.
Moreover, the reductions in~\cite{Jamroga18por,Jamroga21paradoxes-kr,Jamroga20POR-JAIR} are \emph{not} correct if we allow for nested modalities.
The question is: can we adapt the scheme so that it works if we only allow to nest \emph{epistemic} operators?

In this section, we prove that the answer is affirmative.
We also show that the reduction is correct for the subjective semantics of ability, whereas the previous works used the less intuitive (and less popular) objective semantics.

\subsection{Conceptual Machinery}

We first recall the concept of stuttering equivalence.
Intuitively, two paths are stuttering equivalent if they can be divided into corresponding finite segments,
each satisfying exactly the same propositions.
If all states in corresponding segments are also indistinguishable for agents $i \in J$,
then we say the paths are $J-$ stuttering equivalent.

\begin{definition}[($J$-)stuttering equivalence]
\label{def-ste}
Paths $\seq, \seq' \in \Pi_\model(\state)$ are \emph{stuttering equivalent},
denoted $\seq \equiv_{s} \seq'$,
if there exists a partition $B_0 = (\seq[0],\dots,\seq[i_1-1]),\ B_1=(\seq[i_1],\dots,\seq[i_2-1]),\ \ldots$\ of the states of $\seq$,
and an analogous partition $B'_0, B'_1, \ldots$ of the states of $\seq'$,
such that for each $j \geq 0:$ $B_j$ and $B'_j$ are nonempty and finite, and
$V(\state)\cap\hatPV = V(\state')\cap\hatPV$ for every $\state\in B_j$ and $\state'\in B'_j$.

\smallskip
If $\pi \equiv_s \pi'$, and additionally it holds that $\forall j > 0 \quad \forall \state \in B_j, \state' \in B'_j$ :\quad $\state \sim_J \state'$,
then paths $\pi$ and $\pi'$ are \emph{$J$-stuttering equivalent}, denoted $\pi \equiv^J_s \pi'$.

\smallskip
States $\state$ and $\state'$ are \emph{stuttering path equivalent} (resp. \emph{$J$-stuttering path equivalent}),
denoted $\state \equiv_{s} \state'$ (resp. $\state \equiv^J_{s} \state'$),
iff for every path $\pi$ starting from $\state$, there is a path $\pi'$ starting from $\state'$ such that $\pi' \equiv_s \pi$ (resp. $\pi' \equiv^J_s \pi$),
and for every path $\pi'$ starting from $\state'$, there is a path $\pi$ starting from $\state$ such that $\pi \equiv_s \pi'$ (resp. $\pi \equiv^J_s \pi'$).

\smallskip
Models $\model$ and $\model'\subseteq \model$ are \emph{stuttering path equivalent} (resp. \emph{$J$-stuttering path equivalent}),
denoted $\model \equiv_{s} \model'$ (resp. $\model \equiv^J_{s} \model'$),
iff they have the same initial states,
and for each initial state $\iota_i \in \initial$ and each path $\seq \in \Pi_{\model}(\iota_i)$,
there is a path $\seq' \in \Pi_{\model'}(\iota_i)$ such that $\seq \equiv_{s} \seq'$ (resp. $\seq \equiv^J_{s} \seq'$).
\end{definition}

The POR algorithm uses the notions of invisible and independent events.
Intuitively, an event is invisible iff it does not change the valuations of the propositions.\footnote{
  This technical concept of invisibility is not connected to any agent's view, unlike in~\cite{MalvoneMS17}. }
Two events are independent iff at least one is invisible and they are not in the same agent's repertoire.
We designate a subset of agents $A \subseteq \A$ whose events are visible by definition.

\begin{definition}[Invisibility and independence of events]
Let $\model = \IISEps(\AMAS,\initial)$, and $A\subseteq\A$.
An event $\evt \in \events$ is {\em invisible} wrt.~$A$ and $\hatPV$
if $Agent(\evt)\cap A = \emptyset$ and for each two global states $\state, \state' \in \States$
we have that $\state \trans \evt \state'$ implies $V(\state)\cap\hatPV = V(\state')\cap\hatPV$.
The set of all invisible events for $A,\hatPV$ is denoted by $Invis_{A,\hatPV}$,
and its closure, i.e. the set of visible events, by $Vis_{A,\hatPV} = \events \setminus Invis_{A,\hatPV}$.

\smallskip
The notion of \emph{independence} $Ind_{A,\hatPV}\subseteq \events\times \events$ is defined as:
$Ind_{A,\hatPV} = \{(\evt,\evt') \in \events \times \events \mid Agent(\evt) \cap Agent(\evt') = \emptyset\}\ \setminus\ (Vis_{A,\hatPV} \times Vis_{A,\hatPV})$.
Events $\evt, \evt' \in \events$ are called {\em dependent} if $(\evt,\evt') \not \in Ind_{A,\hatPV}$.
If it is clear from the context, we omit the subscript $\hatPV$.\footnote{
The sets of agents' local propositions $\hatPV_i$ are explicitly disjoint in our model (cf. Definition \ref{def:amas}),
allowing for simpler checking of event independence in the actual implementation of POR.}
Note that $\epsilon$ events are always invisible and independent from others,
since they do not modify propositions and we have that $Agent(\epsilon) = \emptyset$ (by Definition \ref{def:undeadlockedIIS}).
\end{definition}

The reduced model (or \emph{submodel}) $\model' \subseteq \model$ obtained with POR extends the same AMAS $\AMAS$ as $\model = \IISEps(\AMAS,\initial)$.
In particular, we have $\States' \subseteq \States$, $\initial = \initial'$, $T$ is an extension of $T'$, and $V' = V|_{\States'}$.
Note that, for each $\state \in \States'$, it holds that $\Pi_{\model'}(\state) \subseteq \Pi_\model(\state)$.

$\model'$ is generated by modifying the standard DFS~\cite{GKPP99},
so that for each $\state$, the successor state $\state_1$ such that $\state \stackrel{\evt}{\rightarrow} \state_1$
is selected from $\ampleset(\state)\cup\set{\epsilon}$ such that $\ampleset(\state) \subseteq enabled(\state)\setminus\set{\epsilon}$.
That is, the algorithm always selects $\epsilon$, plus a subset of the enabled events at $\state$.
This modified DFS is called for each initial state of the model, and we have $\Pi_{\model'} = \bigcup_{\state\in\initial} \Pi_{\model'}(\state)$.
The conditions on the heuristic selection of $\ampleset(\state)$ given below are inspired by~\cite{cgp99,Jamroga20POR-JAIR,peled-on_the_fly}.
\begin{description}
\item[{\bf C1}]
    Along each path $\pi$ in $\model$ that starts at $\state$, each event that is dependent
	on an event in $\ampleset(\state)$ cannot be executed in $\pi$ without an event in $\ampleset(\state)$ being executed first in $\pi$.
	Formally, $\forall \pi \in \Pi_{\model}(\state)$ such that $\pi = \state_0\evt_0\state_1\evt_1\ldots$ with $\state_0 = \state$,
	and $\forall b \in \events$ such that $(b,c) \notin Ind_A$ for some $c \in \ampleset(\state)$,
	if $\evt_i = b$ for some $i \geq 0$, then $\evt_j \in \ampleset(\state)$ for some $j < i$.
\item[{\bf C2}]
    If $\ampleset(\state) \neq enabled(\state)\setminus\set{\epsilon}$, then $\ampleset(\state) \subseteq Invis_A$.
\item[{\bf C3}]
    For every cycle in $\model'$ containing no $\epsilon$-transitions, there is at least one node $\state$ in the cycle for which $\ampleset(\state) = enabled(\state)\setminus\set{\epsilon}$,
		i.e., all the successors of $\state$ are expanded.
\end{description}

Submodel $\model' \subseteq \model$ generated with this algorithm satisfies property $\AE_A$ \cite{Jamroga21paradoxes-kr}:
$\forall\:\strat_A\!\in\!\Sigma_A^\ir\: \quad \forall\:\iota_i\!\in\!\initial\: \quad \forall\:\seq\! \in\! \outcome^{x}_{\model}(\iota_i,\strat_A)\: \quad
 \exists\:\seq'\! \in\! \outcome^{x}_{\model'}(\iota_i,\strat_A)\: \colon\quad \seq\! \equiv_s\! \seq'$,\\
where $x\in\set{\Std,\React}$.

\subsection{POR for \sATLKs}
We will show that the reduction algorithm for \sATLs \cite{Jamroga20POR-JAIR,Jamroga21paradoxes-kr}
can be applied also to formulas that include the knowledge operator (in subformulas of the form $\K_i\varphi$),
provided that $J \subseteq A$.
That is, any agents in these epistemic subformulas are added to the set $A \subseteq \A$ that parametrises the relations of invisibility and independence.

\begin{theorem}\label{por-atlk}
Let $\AMAS$ be an AMAS, $J \subseteq A \subseteq \A$, $\model = \IISEps(\AMAS,\initial)$,
and let $\model' \subseteq \model$ be the reduced model generated by
DFS with the choice of $\ampleset(\state')$ for $\state' \in \States'$ given by conditions {\bf C1-C3}.
Then, for any starting state $\iota_i \in \initial$ and any \sATLKs
formula $\varphi$ over $\hatPV$ that refers only to coalitions $\hat{A}\subseteq A$,
we have that  $\model,\iota_i \satisf[\Y] \varphi$\quad iff \quad $\model',\iota_i \satisf[\Y] \varphi$.
\end{theorem}
\begin{proof}
First, note that conditions {\bf C1-C3} remain unchanged from the reduction algorithm for \sATLs \cite{Jamroga21paradoxes-kr}.
Thus, by \cite[Theorems A.8 and A.9]{Jamroga21paradoxes-kr},\footnote{
The conference paper \cite{Jamroga21paradoxes-kr} does not include a technical
appendix with proofs of Theorems A.8 and A.9, so we refer to its extended arXiv manuscript here.
}
we have that:
\begin{description}
\item[(*)]$\model$ and $\model'$ are stuttering path equivalent.
For each path $\pi = \state_0\evt_0\state_1\evt_1\ldots$ with $\state_0 = \iota_i$ in $\model$,
there is a stuttering equivalent path $\pi' = \state'_0\evt'_0\state'_1\evt'_1\ldots$ with $\state'_0 = \iota_i$ in $\model'$ such that
$Evt(\pi)|_{Vis_A} = Evt(\pi')|_{Vis_A}$,
i.e., $\pi$ and $\pi'$ have the same maximal sequence of visible events for $A$.
\item[(**)]$\model$ and $\model'$ satisfy structural condition $\AE_A$.
\end{description}
That is, we have $\model,\iota_i \satisf[\Y] \varphi$\quad iff \quad $\model',\iota_i \satisf[\Y] \varphi$ for all non-epistemic $\varphi$.
To extend the reasoning to any \sATLKs formula, we first show that the full and reduced model are also $J$-stuttering equivalent,
which then allows to prove that epistemic subformulas are preserved in the reduced model $\model'$.
Finally, we show that these subformulas can be replaced with equivalent new propositions,
effectively reducing the problem to the previously proven case for \sATLs.
\smallskip

Because $J \subseteq A$ (and so all transitions of the agents in group $J$ are visible),
it follows directly from {\bf C2} that if $E(\state) \neq enabled(\state)\setminus\set{\epsilon}$,
then $Agent(\alpha) \cap J = \emptyset$ for any event $\alpha \in \ampleset(\state)$.
This is a direct analogue of the extra condition {\bf CJ} from \cite{lomuscio10partialOrder}.
Together with (*), this implies that the full and reduced model are also $J$-stuttering equivalent:
\begin{description2}
\item[(***)]$\model \equiv^J_s \model'$.
\end{description2}

Consider any subformula $\varphi = \K_i \psi$.
As per the syntax of \sATLKs, temporal operators and strategic modalities cannot be nested inside $K_i$,
so $\varphi$ is a purely epistemic formula that only contains knowledge operator(s) and propositional variables with Boolean connectives.
Now, we will show it follows from (***) that epistemic subformulas are preserved in the reduced model, i.e.,
for any state $\state$ such that $\state\equiv^J_s\state'$, we have $\model,\state \satisf[\Y] \varphi$ iff $\model',\state' \satisf[\Y] \varphi$:

\smallskip
($\Rightarrow$) Assume that $\model,\state \satisf[\Y] \K_i \psi$.
Let $\States_\psi = \{\state_\psi \in \States \mid \state \sim_i \state_\psi \}$, and take $\state'_\psi$ such that $\state' \sim_i \state'_\psi$.
We need to show that $\model',\state'_\psi \satisf[\Y] \psi$.
From $\state \equiv^J_s \state'$ and by transitivity of relation $\sim_i$, we have that $\state'_\psi \in \States_\psi$.
So, clearly $\model,\state'_\psi \satisf[\Y] \psi$.
As $\state_\psi \equiv^J_s \state'_\psi$, it follows from the inductive assumption that $\model',\state'_\psi \satisf[\Y] \psi$.
Hence, $\model',\state' \satisf[\Y] \K_i \psi$.

\smallskip
($\Leftarrow$) Assume that $\model',\state' \satisf[\Y] \K_i \psi$.
Let $\States'_\psi = \{\state'_\psi \in \States' \mid \state' \sim_i \state'_\psi \}$, and take $\state_\psi$ such that $\state \sim_i \state_\psi$.
We need to show that $\model,\state_\psi \satisf[\Y] \psi$.
Consider a path $\pi \in \model$ that contains $\state_\psi$.
From (***), there is a path $\pi' \in \model'$, which contains a state $\state''_\psi \in \States'$, such that $\state_\psi \equiv^J_s \state''_\psi$.
By transitivity of $\sim_i$, we get that $\state''_\psi \in \States'_\psi$, and thus $\model',\state''_\psi \satisf[\Y] \psi$.
As $\state_\psi \equiv^J_s \state''_\psi$, it follows from the inductive assumption that $\model,\state_\psi \satisf[\Y] \psi$.
Hence, $\model,\state \satisf[\Y] \K_i \psi$.

\smallskip
From the above we get that any epistemic subformula $K_i \psi$ holds in the reduced model $\model'$ iff
it holds in the corresponding state of the ($J$-stuttering equivalent) full model $\model$.
Now, we introduce auxiliary propositional variables to replace epistemic subformulas, including nested ones.
\smallskip

Consider subformulas $\varphi_0,\varphi_1,\dots$, where $\varphi_0 = \varphi$, and for all $i > 0$, $\varphi_i$
is an epistemic subformula nested in $\varphi_{i-1}$.
Note that in the reduced model $\model'$, one can add a set of new propositional variables $\PV' = \bigcup_i \{ \prop{sat_{\varphi_i}} \}$ to $\PV$,
and extend the valuation function accordingly, so that we have $V': \States' \rightarrow 2^{\PV \cup \PV'}$,
and $\prop{sat_{\varphi_i}}$ is true in state $\state' \in \States'$ iff $\model',\state' \satisf[\Y] \varphi_i$.
That is, for each epistemic (sub)formula $\varphi_i$, a new proposition $\prop{sat_{\varphi_i}}$ is added,
whose valuation in each state $\state' \in \States'$ corresponds to the satisfaction of $\varphi_i$ in that state of $\model'$.
Then, for the formula $\varphi' = \prop{sat_{\varphi_0}}$, it clearly holds that $\model',\state' \satisf[\Y] \varphi'$ iff $\model',\state' \satisf[\Y] \varphi$.
Furthermore, since epistemic subformulas only refer to agents $i \in J$ and we have that $J \subseteq A$,
it follows that $Vis_{A,\PV} = Vis_{A,\PV\cup\PV'}$ and $Ind_{A,\PV} = Ind_{A,\PV\cup\PV'}$.
That is, replacing epistemic subformulas with new propositions in this manner does not affect the visibility or independence of events wrt. $A$ and $\PV$.
Hence, we also have $\model',\state' \satisf[\Y] \varphi'$ iff $\model,\state \satisf[\Y] \varphi$,
from (*) and (**) and by \cite[Theorems A.8 and A.9]{Jamroga21paradoxes-kr}, as $\varphi'$ is a \sATLs formula.
But from the construction of $\varphi'$, we have $\model',\state' \satisf[\Y] \varphi'$ iff $\model',\state' \satisf[\Y] \varphi$,
so it also holds that $\model,\state \satisf[\Y] \varphi$ iff $\model',\state' \satisf[\Y] \varphi$ for any \sATLKs formula $\varphi$,
both for standard outcomes \cite[Theorem A.8]{Jamroga21paradoxes-kr} and under the opponent reactivity condition \cite[Theorem A.9]{Jamroga21paradoxes-kr}.
Thus, in particular we have that $\model,\iota_i \satisf[\Y] \varphi$\quad iff \quad $\model',\iota_i \satisf[\Y] \varphi$, QED.
\end{proof}

\subsection{POR for Subjective Semantics of Ability}

Recall the subjective semantics of strategic ability in Section~\ref{sec:atl}.
We will show that the reduction scheme remains applicable in this setting,
i.e., when the set of initial states is comprised of previously designated subset $\initial \subseteq \States$,
plus all states indistinguishable from those in $\initial$ for coalition agents.

\begin{theorem}\label{por-atlk-subjective}
Let $x\in\set{\Std,\React}$, $\hat{A}\subseteq A$ and
$\initial_{\hat{A}} = \initial \cup \set{ \state \in \States \mid \exists_{\iota_i \in \initial} \exists_{j\in\hat{A}} : \state \sim_j \iota_i }$.
Let $\model = \fullmodel(\AMAS,\initial_{\hat{A}})$,
and let $\model'$ be the reduction of $\model$ generated by POR using conditions \textbf{C1-C3} for the choice of ample sets.
Then, for any initial state $\iota_i \in \initial_{\hat{A}}$ and any
\sATLKs formula $\varphi$ that refers only to coalition $\hat{A}$,
we have that $\model,\iota_i \satisfS^{x} \varphi$\quad iff \quad $\model',\iota_i \satisfS^{x} \varphi$.
\end{theorem}
\begin{proof}[Proof sketch]
For each $\iota_i \in \initial_{\hat{A}}$,
take $\model_i$ constructed by DFS starting from $\iota_i$ (i.e., with a single initial state),
and let $\model'_i$ be the reduction of $\model_i$ generated by POR.

Take any path $\pi \in \Pi_\model$. Clearly, $\pi \in \Pi_{\model_i}$ for some $i > 0$.
By Theorem~\ref{por-atlk}, we have $\model_i \equiv^J_s \model'_i$, so there is a $J$-stuttering equivalent path $\pi' \in \Pi_{\model'_i}$.
From the construction by DFS, $\Pi_{\model'} = \bigcup_{i \in \initial_{\hat{A}}} \Pi_{\model'_i}$.
Hence, $\pi' \in \Pi_{\model'}$, which implies that $\model \equiv^J_s \model'$. (*)

Take any joint strategy $\strat_{\hat{A}}$.
The subjective outcome of $\strat_{\hat{A}}$ in $\model$ (resp. $\model'$)
is the sum of objective outcomes of $\strat_{\hat{A}}$ in $\model_i$ (resp. $\model'_i$).
But from $\AE_A$, $\outcome^{x}_{\model_i}(\iota_i,\strat_{\hat{A}}) \equiv_s \outcome^{x}_{\model_i'}(\iota_i,\strat_{\hat{A}})$.
So, analogously to the reasoning for (*), it follows that
$\bigcup_{i \in \initial_{\hat{A}}} \outcome^{x}_{\model_i}(\iota_i,\strat_{\hat{A}}) \equiv_s \bigcup_{i \in \initial_{\hat{A}}} \outcome^{x}_{\model_i'}(\iota_i,\strat_{\hat{A}})$.
Hence, $\model$ and $\model'$ also satisfy $\AE_A$. (**)

Since (*) and (**), the thesis follows from Theorem~\ref{por-atlk}, as in the case for objective semantics of strategic ability.
\end{proof}

\section{Experiments and Results}\label{sec:experiments}

\begin{table}[t]
\begin{adjustbox}{width=\resultsScale\textwidth,center}
    \begin{tabular}{|cHHH|c|c|c|c|c|c|c|c|c|c|c|c|}
        \hline
        \multirow{2}{*}{\#A} & \multirow{2}{*}{\#V} & \multirow{2}{*}{\#CV} & \multirow{2}{*}{\#C}    & \multirow{2}{*}{\#R}   & \multicolumn{5}{c|}{Full Model}                         & \multicolumn{5}{c|}{Reduced Model}                & \multirow{2}{*}{Result} \\ \cline{6-15}
                              &                      &                       &                         &                        & \#st   & \#tr   & Seq.                & Par.  & Appr.   & \#st     & \#tr   & Seq.     & Par.     & Appr.   &                         \\ \hline
        4                     & 1                    & 1                     & 3                       & 3                      & 3.63e4 & 7.46e4 & 0.003               & 0.009 & 1.121   & 2.60e4   & 5.99e4 & 0.001    & 0.002    & 0.184   & True                    \\ \hline
        4                     & 1                    & 1                     & 3                       & 5                      & 5.62e4 & 1.15e5 & 0.004               & 0.003 & 0.345   & 4.01e4   & 9.26e4 & 0.002    & 0.002    & 0.283   & True                    \\ \hline
        4                     & 1                    & 1                     & 3                       & 10                     & 1.06e5 & 2.18e5 & 0.009               & 0.005 & 0.691   & 7.55e4   & 1.74e5 & 0.004    & 0.002    & 0.563   & True                    \\ \hline
        5                     & 2                    & 1                     & 3                       & 3                      & 1.55e6 & 5.91e6 & 0.158               & 0.004 & 14.78   & 1.09e6   & 4.65e6 & 0.112    & 0.021    & 12.99   & True                    \\ \hline
        6                     & 3                    & 1                     & 3                       & 3                      & 7.61e7 & 4.98e8 & 0.524               & 0.051 & 41.24   & 5.34e7   & 3.82e8 & 0.427    & 0.042    & 37.35   & True                    \\ \hline
        7                     & 4                    & 1                     & 3                       & 3                      & \multicolumn{5}{c|}{model generation timeout}                            & \multicolumn{5}{c|}{model generation timeout}                      & -                       \\ \hline
        \end{tabular}
\end{adjustbox}
\caption{Verification of $\varphi_{vuln,i,k}$ for the first candidate ($i=1$) and $k=\#R$ revotes}
\label{tab:res1}
\end{table}

\begin{table}[t]
\begin{adjustbox}{width=\resultsScale\textwidth,center}
    \begin{tabular}{|cHHH|c|c|c|c|c|c|c|c|c|c|c|c|}
        \hline
        \multirow{2}{*}{\#Ag} & \multirow{2}{*}{\#V} & \multirow{2}{*}{\#CV} & \multirow{2}{*}{\#C}    & \multirow{2}{*}{\#R}   & \multicolumn{5}{c|}{Full Model}                         & \multicolumn{5}{c|}{Reduced Model}                & \multirow{2}{*}{Result} \\ \cline{6-15}
                              &                      &                       &                         &                        & \#st   & \#tr   & Seq.                & Par.  & Appr.   & \#st     & \#tr   & Seq.     & Par.     & Appr.   &                         \\ \hline
        4                     & 1                    & 1                     & 3                       & 3                      & 3.63e4 & 7.46e4 & 0.003               & 0.010 & 1.103   & 2.60e4   & 5.99e4 & 0.002    & 0.003    & 0.166   & True                    \\ \hline
        4                     & 1                    & 1                     & 3                       & 5                      & 5.62e4 & 1.15e5 & 0.004               & 0.005 & 0.348   & 4.01e4   & 9.26e4 & 0.003    & 0.003    & 0.280   & True                    \\ \hline
        4                     & 1                    & 1                     & 3                       & 10                     & 1.06e5 & 2.18e5 & 0.008               & 0.009 & 0.700   & 7.55e4   & 1.74e5 & 0.005    & 0.004    & 0.567   & True                    \\ \hline
        5                     & 2                    & 1                     & 3                       & 3                      & 1.55e6 & 5.91e6 & 0.160               & 0.055 & 14.03   & 1.09e6   & 4.65e6 & 0.112    & 0.053    & 12.49   & True                    \\ \hline
        6                     & 3                    & 1                     & 3                       & 3                      & 7.61e7 & 4.98e8 & 0.602               & 0.083 & 42.44   & 5.34e7   & 3.82e8 & 0.501    & 0.057    & 38.20   & True                    \\ \hline
        7                     & 4                    & 1                     & 3                       & 3                      & \multicolumn{5}{c|}{model generation timeout}                            & \multicolumn{5}{c|}{model generation timeout}                      & -                       \\ \hline
        \end{tabular}
\end{adjustbox}
\caption{Verification of $\varphi_{vuln,i,k}$ for the last candidate ($i=\#C$) and $k=\#R$ revotes}
\label{tab:res2}
\end{table}

\begin{table}[t]
\begin{adjustbox}{width=\resultsScale\textwidth,center}
    \begin{tabular}{|cHHH|c|c|c|c|c|c|c|c|c|c|c|c|}
        \hline
        \multirow{2}{*}{\#Ag} & \multirow{2}{*}{\#V} & \multirow{2}{*}{\#CV} & \multirow{2}{*}{\#C}    & \multirow{2}{*}{\#R}   & \multicolumn{5}{c|}{Full Model}                         & \multicolumn{5}{c|}{Reduced Model}                & \multirow{2}{*}{Result} \\ \cline{6-15}
                              &                      &                       &                         &                        & \#st   & \#tr   & Seq.                & Par.  & Appr.   & \#st     & \#tr   & Seq.     & Par.     & Appr.   &                         \\ \hline
        4                     & 1                    & 1                     & 3                       & 3                      & 3.63e4 & 7.46e4 & 0.303               & 0.317 & 1.128   & 2.60e4   & 5.99e4 & 0.202    & 0.205    & 0.179   & False                   \\ \hline
        4                     & 1                    & 1                     & 3                       & 5                      & 5.62e4 & 1.15e5 & 0.524               & 0.592 & 0.325   & 4.01e4   & 9.26e4 & 0.411    & 0.503    & 0.280   & False                    \\ \hline
        4                     & 1                    & 1                     & 3                       & 10                     & 1.06e5 & 2.18e5 & 0.721               & 0.668 & 0.459   & 7.55e4   & 1.74e5 & 0.525    & 0.512    & 0.364   & False                    \\ \hline
        5                     & 2                    & 1                     & 3                       & 3                      & 1.55e6 & 5.91e6 & 2.146               & 1.257 & 0.981   & 1.09e6   & 4.65e6 & 1.513    & 1.003    & 0.583   & False                    \\ \hline
        6                     & 3                    & 1                     & 3                       & 3                      & 7.61e7 & 4.98e8 & 5.232               & 3.228 & 1.892   & 5.34e7   & 3.82e8 & 4.986    & 2.427    & 1.092   & False                    \\ \hline
        7                     & 4                    & 1                     & 3                       & 3                      & \multicolumn{5}{c|}{model generation timeout}                            & \multicolumn{5}{c|}{model generation timeout}                      & -                       \\ \hline
        \end{tabular}
\end{adjustbox}
\caption{Verification of $\varphi_{vuln,i,k}$ for the first candidate ($i=1$) and $k=\#R-1$ revotes}
\label{tab:res3}
\end{table}

\begin{table}[t]
\begin{adjustbox}{width=\resultsScale\textwidth,center}
    \begin{tabular}{|cHHH|c|c|c|c|c|c|c|c|c|c|c|c|}
        \hline
        \multirow{2}{*}{\#Ag} & \multirow{2}{*}{\#V} & \multirow{2}{*}{\#CV} & \multirow{2}{*}{\#C}    & \multirow{2}{*}{\#R}   & \multicolumn{5}{c|}{Full Model}                         & \multicolumn{5}{c|}{Reduced Model}                & \multirow{2}{*}{Result} \\ \cline{6-15}
                              &                      &                       &                         &                        & \#st   & \#tr   & Seq.                & Par.  & Appr.   & \#st     & \#tr   & Seq.     & Par.     & Appr.   &                         \\ \hline
        4                     & 1                    & 1                     & 3                       & 3                      & 3.63e4 & 7.46e4 & 0.302               & 0.311 & 0.180   & 2.60e4   & 5.99e4 & 0.201    & 0.213    & 0.126   & False                   \\ \hline
        4                     & 1                    & 1                     & 3                       & 5                      & 5.62e4 & 1.15e5 & 0.519               & 0.584 & 0.310   & 4.01e4   & 9.26e4 & 0.410    & 0.475    & 0.283   & False                    \\ \hline
        4                     & 1                    & 1                     & 3                       & 10                     & 1.06e5 & 2.18e5 & 0.742               & 0.627 & 0.462   & 7.55e4   & 1.74e5 & 0.558    & 0.544    & 0.370   & False                    \\ \hline
        5                     & 2                    & 1                     & 3                       & 3                      & 1.55e6 & 5.91e6 & 2.160               & 1.358 & 0.942   & 1.09e6   & 4.65e6 & 1.621    & 1.009    & 0.519   & False                    \\ \hline
        6                     & 3                    & 1                     & 3                       & 3                      & 7.61e7 & 4.98e8 & 5.504               & 3.516 & 1.903   & 5.34e7   & 3.82e8 & 5.110    & 2.380    & 1.112   & False                    \\ \hline
        7                     & 4                    & 1                     & 3                       & 3                      & \multicolumn{5}{c|}{model generation timeout}                            & \multicolumn{5}{c|}{model generation timeout}                      & -                       \\ \hline
        \end{tabular}
\end{adjustbox}
\caption{Verification of $\varphi_{vuln,i,k}$ for the last candidate ($i=\#C$) and $k=\#R-1$ revotes}
\label{tab:res4}
\end{table}

\begin{figure}[t]
    \begin{adjustbox}{width=0.35\textwidth,center}
        \begin{tikzpicture}  
          
          \begin{axis}  
          [  
              ybar,  
              enlargelimits=0.15,  
              ylabel={\ Time in seconds}, 
              xlabel={\#Agents},  
              legend style={at={(0.1,0.8)},
              anchor=west,legend columns=1}, 
              symbolic x coords={4, 5, 6}, 
              xtick=data,  
              ymode=log,
              nodes near coords align={vertical},  
              ]  
          \addplot[fill = black!10] coordinates {(4,4) (5,347) (6,32548) };  
          \addplot[fill = black!70] coordinates {(4,17) (5,728) (6,66102) };  

          \legend{Full Model, Reduced Model}
            
          \end{axis}  
        \end{tikzpicture}  
    \end{adjustbox}
    \caption{Model generation times for 3 candidates 3 revotes}
    \label{fig:mtime}
\end{figure}
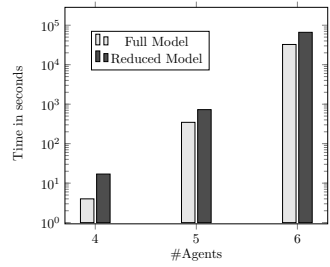

In this section, we give a brief summary of the experimental results.

\para{Models and formulas.}
The scalable class of models has been described in detail in
Section~\ref{sec:specification}.
They can be configured using four parameters:
the number of voters ($V$), the number of coerced voters ($CV$), the number of candidates ($C$) and the number of revoting turns ($R$).
In all experiments we use configurations with one coerced voter and three candidates.
We also use the \emph{coercion-vulnerability formula} 
\begin{multline*}
\varphi_{vuln,i,k} \equiv\\ \coop{\Coercer}\Always ((\prop{end} \land \prop{revote_{v1}=k} \land \prop{voted_{v1}=i}) \then \K_{Coercer}\prop{voted_{v1}=i})
\end{multline*}
saying that the coercer has a strategy so that, at the end of the election, if the voter has effectively voted for candidate $i$, then the coercer knows about it.
This can be seen as the opposite of \emph{receipt-freeness} and \emph{coercion-resistance} formalizations in~\cite{Juels05coercion,Tabatabaei16expressing}, see Section~\ref{sec:properties} for more details.

\para{Configuration of the experiments.}
The test platform was a server equipped with ninety-six 2.40 GHz Intel Xeon Platinum 8260 CPUs, 991 GB RAM, and 64-bit Linux. All times are given in seconds.
The algorithms have been implemented in C++.
The code is available on github: \href{https://github.com/blackbat13/ATLFormulaCheckerC}{github.com/blackbat13/ATLFormulaCheckerC}.

\para{Results.}
We present the verification results in the Tables~\ref{tab:res1}-\ref{tab:res4} and model generation times in the Figure~\ref{fig:mtime}.
All times are given in seconds and the timeout for verification was set to 1 hour and 20 hours for model generation.
Model generation times are presented using a logarithmic scale.
We present the size of the model and the verification times both for the full and the reduced model.
We have compared time results for four verification methods, as described in Section~\ref{sec:verification}: the sequential strategy synthesis (Seq.), the parallelized version (Par.) and the fixpoint approximation (Appr.).

\para{Results: \Domino and flexible distributed algorithm.}
We omit the results for the \Domino algorithm, as it performed much slower than other algorithms.
The reason is that in this scenario there is no redundancy in the strategy space, and hence no room for elimination of dominated strategies. In consequence, no gain is achieved compared to the standard DFS strategy synthesis.

The same applies to the flexible distributed algorithm proposed in Section~\ref{sec:distr-flex}. Apparently,
cloning the model and synchronization between the threads grossly outweighs the benefits of using parallel computation.

\para{Discussion of the results.}
As the results show, the simple parallel verification performs quite well in most cases.
Only for relatively small models the sequential algorithm achieves better performance than the parallel algorithm, which is the result of a overhead associated with the parallelization.
As various experiments have shown, the performance of the parallel algorithm is heavily dependent on the structure of the model.
Depending on the amount of branching and loops in the model, it can result in a timeout even for small models.
The reason for this behavior seems simple: the space of all strategies is too big to manage.
As the strategy is generated from top to bottom, less branching means less configurations to check.

The fixpoint approximation algorithm performs well in cases where no strategy can be found, as it quickly reaches the fixpoint.
On the other hand, when the formula is satisfied, multiple loops maybe required before reaching the fixpoint, with each loop adding more states to the computation set.

\para{Challenges and lessons learned.}
When conducting the experiments, we have encountered two main difficulties: high memory usage and slow model generation.
The first one results in a memout for ordinary computers.
The second one results in a timeout for more powerful machines with hundreds of gigabytes of RAM.
Before moving to better computers, we have tried to overcome the memory usage problem by implementing a communication with an external database engine.
The idea was to store parts of the model during the generation in the database.
That, in theory, allows to shift some part of the memory requirements to other parts of the system.
Unfortunately, this also heavily impacted generation times, resulting in approximately 10 times slower computation, which lead us to abandoning this idea.

To overcome the high model generation times one can also try to parallelize the generation of models.
Our experiments have shown that the parallel algorithm, if implemented in a clever way, can greatly reduce the computation times.
The model generation procedure works similarly to the DFS algorithm, which suggests that its parallel version can be as effective.

\section{Conclusions}\label{sec:conclusions}
Modal logics for MAS, and the related verification problems, have been studied for many years. Unfortunately, they are characterized by high computational complexity, both in theory and in practice.
On the other hand, it is necessary to try and verify real-life scenarios, with all their complexity, to make substantial progress in the field.

In this paper, we propose a hands-on case study in verification of a genuine protocol for secure voting.
We use the ``all out'' approach, implementing multiple existing techniques (depth-first strategy search, domination-based strategy synthesis, fixpoint approximation) as well as proposing novel adaptations of others (partial-order reduction, distributed strategy search).
Of those, partial order reduction, simple DFS, simple distributed DFS, and fixpoint approximation show very promising performance. Moreover, they produce significant gains in some of the tested configurations.
On the other hand, the domination-based synthesis and flexible distributed turn out rather ill-fitted to the model checking task at hand.
Overall, the experimental results are promising, and suggest that the time is ripe for the community to engage more in realistic application of the algorithms that we develop.

We also emphasize that the extension of partial order reduction, presented here, is a nontrivial technical result in its own right, as it is the first POR algorithm that handles the nesting of epistemic modalities in the context of strategic ability.

\bibliographystyle{aiml22}

\end{document}